\documentclass[twocolumn]{aa}
\usepackage{graphicx}
\include{epsf}
\usepackage{txfonts}
\begin{document}
   \title{A low upper-limit on the lithium isotope ratio in HD140283\thanks{Based on data collected
          at the Subaru Telescope, which is operated by the National
          Astronomical Observatory of Japan}}

%   \subtitle{
%   Based on data collected at the Subaru Telescope, which is operated by the
%   National Astronomical Observatory of Japan}

   \author{Wako Aoki\inst{1}
	  Susumu Inoue\inst{2}
	  Satoshi Kawanomoto\inst{1}
	  Sean G. Ryan\inst{3}
	  Iain M. Smith\inst{3}
          Takeru K. Suzuki\inst{1,4}
          \and
          Masahide Takada-Hidai\inst{5}}

   \offprints{S. G. Ryan}

   \institute{
   National Astronomical Observatory,  2-21-1 Osawa, Mitaka, Tokyo 181-8588, Japan.\\
   \email{aoki.wako@nao.ac.jp kawanomo@optik.mtk.nao.ac.jp stakeru@th.nao.ac.jp}
   \and
   Max-Planck-Institut f\"ur Astrophysik, Karl-Schwarzschild-Str. 1, D-85741 Garching, Germany.\\
   \email{inouemu@mpa-garching.mpg.de}
   \and
   Centre for Earth, Planetary, Space and Astronomical Research,
   Dept of Physics \& Astronomy, The Open University, Walton Hall,
   Milton Keynes, MK7 6AA, UK.\\
   \email{s.g.ryan@open.ac.uk }
   \and
   Department of Astronomy, School of Science, University of Tokyo,
   7-3-1 Hongo, Bunkyo-ku, Tokyo 113-0033, Japan.\\             
   \and
   Liberal Arts Education Center, Tokai University, 1117 Kitakaname,
   Hiratsuka-shi, Kanagawa, 259-1292, Japan. \\
   \email{hidai@apus.rh.u-tokai.ac,jp}
   }

   \date{Received January xx, 2003; accepted xxx xx, xxxx}
   \authorrunning{W. Aoki et al.}
   \titlerunning{Lithium isotope ratio in HD~140283}

   \abstract{ We have obtained a high-S/N (900-1100),
high-resolving-power (R=95000) spectrum of the metal-poor subgiant
HD~140283 in an effort to measure its $^6$Li/$^7$Li isotope
ratio. From a 1-D atmospheric analysis, we find a value consistent
with zero, $^6$Li/$^7$Li = 0.001, with an upper limit of
$^6$Li/$^7$Li$<$0.026.  This measurement supersedes an earlier
detection (0.040$\pm$0.015($1\sigma$)) by one of the
authors. HD~140283 provides no support for the suggestion that Population II
stars may preserve their $^6$Li on the portion of the subgiant branch
where $^7$Li is preserved. However, this star does not defeat the
suggestion either; being at the cool end of subgiant branch of the
Spite plateau, it may be sufficiently cool that $^6$Li depletion has
already set in, or the star may be sufficiently metal poor that little
Galactic production of $^6$Li had occurred. Continued investigation of
other subgiants is necessary to test the idea.  We also consider the
implications of the HD140283 upper limit in conjunction with other
measurements for models of $^6$Li production by cosmic rays from
supernovae and structure formation shocks.

   }

   \maketitle
%
%________________________________________________________________

\section{Introduction}

Although the Big Bang is believed to be the major producer of the
$^7$Li seen in Population II (Pop. II) stars, it is not believed to be
a significant source of the lighter isotope, $^6$Li. A range of
possible sites exist for $^6$Li including not only spallative and
fusion sources normally associated with supernova-accelerated Galactic
cosmic rays (Walker et al. 1985; Steigman \& Walker 1992), but also
stellar flares (Deliyannis \& Malaney 1995) and possibly shocks
produced by large-scale-structure formation (Suzuki \& Inoue 2002).
Measurements of the $^6$Li/$^7$Li ratio in metal-poor stars could
therefore provide important constraints on Li production following the
Big Bang. The isotope ratio can also constrain possible destruction of
Li, since $^6$Li is destroyed in stars at a lower temperature than
$^7$Li. $^6$Li is more susceptible to destruction than $^7$Li in some
Li-depleting mechanisms (Deliyannis 1990), but not necessarily in
slow-mixing models where the less-fragile elements $^7$Li and $^9$Be
are depleted in concert with one another (Deliyannis et al. 1998).
Consequently, reliable measurements of the Galactic evolution of
$^6$Li as well as $^7$Li could have wide implications. Since the
fusion mechanism (Steigman \& Walker 1992) produces $^6$Li without
co-producing beryllium or boron, observations of Be (Boesgaard et
al. 1999) and B (Duncan et al. 1997) do not adequately constrain Li
production, especially at the earlier epochs where fusion was more
important than spallation. Models of spallative production of Be and B
are also affected by long-running uncertainties in the Galactic
chemical evolution of the most relevant heavy nucleus,
oxygen. Consequently, the yield of $^6$Li/$^9$Be depends not only on
the cosmic ray energy spectrum (which may include a substantial low
energy component (LEC); Vangioni-Flam, Lehoucq, \& Cass\'e\ 1994),
confinement (Prantzos, Cass\'e, \& Vangioni-Flam 1993; Fields, Olive,
\& Schramm 1994), the evolution with metallicity of the flux
(Yoshii, Kajino \& Ryan 1997) and composition (Fields et al. 1994;
Ramaty et al. 1997; Vangioni-Flam et al. 1999), but also on the fast
particle and interstellar medium abundances. Depending on the particle
source abundance distribution, the production ratio of $^6$Li/$^9$Be
ranges over 3--100 (Ramaty, Kozlovsky, \& Lingenfelter 1996).
Consequently, it is necessary to constrain $^6$Li evolution directly
from measurements of the $^6$Li/$^7$Li ratio.

Unfortunately, the measurement of $^6$Li in stellar spectra is very
difficult.  The 6707~{\AA} transition, the only Li feature strong
enough to permit an attempt, is a fine-structure doublet, and the
isotopic displacement of the $^6$Li lines from the $^7$Li lines is
comparable to both the doublet separation and to the intrinsic line
width. The latter is determined primarily by thermal Doppler
broadening in the hot stellar atmosphere and by poorly characterised
non-thermal (turbulent) motions. These factors, combined with the low
fraction ($<$10\%) of $^6$Li, mean that high-resolution,
high-signal-to-noise spectra are required. Even then, stellar models
(Deliyannis 1990), supported by observational evidence (Smith, Lambert
\& Nissen 1993), indicate that $^6$Li is depleted below detection
levels in all but the hottest main-sequence, Pop. II stars. Finding
Pop. II stars hot enough to preserve $^6$Li and bright enough to yield
high S/N at high spectral resolution has been difficult. Progress was
made with 3- to 4-metre telescopes during the last decade: Smith et
al. (1993, 1998) find $^6$Li/Li = 0.06~$\pm$~0.03 in HD~84937 and
$^6$Li/Li = 0.05~$\pm$~0.03 in BD+26$^\circ$3578; Hobbs \& Thorburn
(1994, 1997) measure $^6$Li/$^7$Li = 0.08~$\pm$~0.04 in HD~84937; and
Cayrel et al. (1999) obtain $^6$Li/$^7$Li = 0.052~$\pm$~0.019 in
HD~84937. The reality of the $^{6}$Li detection in HD84937 is
supported by an analysis of its K {\small I} 7698 {\AA} line (Smith et
al. 2001). Also important are the numerous non-detections in other
stars reported in those works. Isotope ratios have also been measured
for a number of metal-poor disc stars (Nissen et al. 1999). The advent
of 8-m telescopes should improve the situation.

Deliyannis (1990, his Fig. 7) noted that $^6$Li survival might also be
high in subgiants containing Spite-plateau Li abundances, at least
down to $\sim$5800 K, since these stars had high effective
temperatures on the main sequence.  Deliyannis \& Ryan (2000) set out
to test this directly, and on the basis of a spectrum analysis by
S.G.R. claimed a detection of $^6$Li/$^7$Li at
$0.040\pm$0.014($1\sigma$) in the subgiant HD~140283. We set out to
verify this observation and analysis using the new High Dispersion
Spectrograph (HDS) on the Subaru 8.2-m telescope (Noguchi et
al. 2002). In the following sections of this paper, we present data
with a S/N around 1000 per pixel, and report our finding of a very low
upper limit, $^6$Li/$^7$Li $<$ 0.018, which supersedes S.G.R.'s
earlier analysis.

\section{Observation}

During commissioning of HDS, we sought a very high-S/N observation of
HD~140283 to verify the earlier $^6$Li detection at the same time as
proving the capabilities of the new spectrograph. The spectrum was
obtained over two nights, 22 and 29 July 2001, from thirteen exposures
totalling 82~minutes.  This gave a S/N = 900-1100 per 0.018 {\AA} pixel
around 6707 {\AA} at $R \equiv \lambda/\Delta\lambda$ = 95000, using a
slit width of 0.4~arcsec (0.2~mm). The seeing was typically
0.55-0.6~arcsec, so the slit was well illuminated. This allows us to
estimate the instrumental profile from the comparison spectrum, which
we do below.  The spectrum was wavelength calibrated using a ThAr
spectrum, and gave typical RMS errors of 0.0015 {\AA}.

\section{ Spectral analysis }

\subsection{General approach}

Our procedure was to compare the observed data with synthetic spectra
computed using code originating with Cottrell \& Norris (1978) and a
1D model atmosphere grid by R. A. Bell (1983, priv. comm.).  A model
with $T_{\rm eff}$ = 5750~K, $\log g$ = 3.4, [Fe/H] = $-$2.5, and
microturbulence $\xi$ = 1.4~km~s$^{-1}$ was used, along with the Li
line list of Smith et al. (1998).  Spectra were synthesised for
various values of $^6$Li/$^7$Li and convolved with macroturbulent and
instrumental profiles.  In order to do this we first determined the
instrumental profile and then found (model-dependent) constraints on
macroturbulence using procedures described in Sects. 3.2 and 3.3.

%log$_{10}$${{g}\over{\rm cm s^{-1}}}$

To determine the best fit when comparing synthetic and observed
spectra, a $\chi^2$ test was employed, where \begin{center} $\chi^2
\equiv \Sigma[{(O_i - S_i)^2 \over \sigma_i^2}]$ \end{center} and
$O_i$ is the observed continuum-normalised flux, $S_i$ is the
synthesised flux, and $\sigma_i$ is the standard deviation of the
observed points defining the continuum.  From this we calculated the
reduced $\chi^2$, $\chi_r^2$, defined as \begin{center} $\chi_r^2
\equiv {\chi^2 \over (\nu - 1)}$ \end {center} where $\nu$ is the
number of degrees of freedom (e.g. Smith et al. 1998).  The best fit
from a set of synthesised spectra is the one that minimises
$\chi_r^2$.

Once the observed spectra were normalised using neighbouring continuum
windows, four variables affect the comparison between spectra: the
abundance $A$($X$) of element $X$, the wavelength shift
$\Delta\lambda$, the macroturbulence $\Gamma$, and the
$^6\mathrm{Li}/^7\mathrm{Li}$ ratio.  To determine these, several
iterations are needed as they are dependent on each other at some
level.

The wavelength shift $\Delta\lambda$ that is allowed for each line may
have any of several possible origins: an error in the applied Doppler
correction for the star's motion; errors in the wavelength calibration
of the ThAr frame (RMS = 0.0015~{\AA}); an error in the wavelength of
the line listed in the spectrum-synthesis linelist, which will differ
from line to line; and possible variations in the motions of different
elements, ionisation states, and excitation states in a dynamic, 3D,
real stellar atmosphere.  We return to these shifts below, in Sect. 3.3.

\subsection{\it Instrumental profile}

The instrumental profile was calculated from a ThAr
hollow-cathode-lamp spectrum over the interval 6660-6730 {\AA} taken
with the same instrumentation setup as the stellar exposures. Nineteen
emission lines of various strengths were isolated and normalised to
the same strength by scaling them to the height of a fitted
Gaussian. (In doing this we are not claiming that the ThAr profiles
are Gaussian, merely that a Gaussian provides a useful reference
profile for normalising their heights.) On closer examination it was
noted that six had weak lines in their wings, significant enough to
impact weakly on our estimate of the instrumental profile, and so were
rejected from the analysis.  The remaining 13 unblended lines were
then overlaid and averaged to give the instrumental profile shown in
Figure~1(a). 
{\bf 
(The instrumental profiles for several setups have been 
given in Noguchi et al. (2002). Our result is similar.)
}
It is comforting that the lines give essentially
identical profiles despite their different intensities.  From the
fall-off of the flux away from the line core, we believe that the
lines effectively end around $\pm$0.3 {\AA}, but we chose to truncate
our calculations at 0.2 {\AA}. We estimate that $<0.5$\% of the flux
lies beyond $\pm$0.2 {\AA}.

\begin{figure}
% li6hds_f1.eps goes here
% \vspace{11cm}
\epsfxsize=084mm
\epsfbox[33 173 295 689]{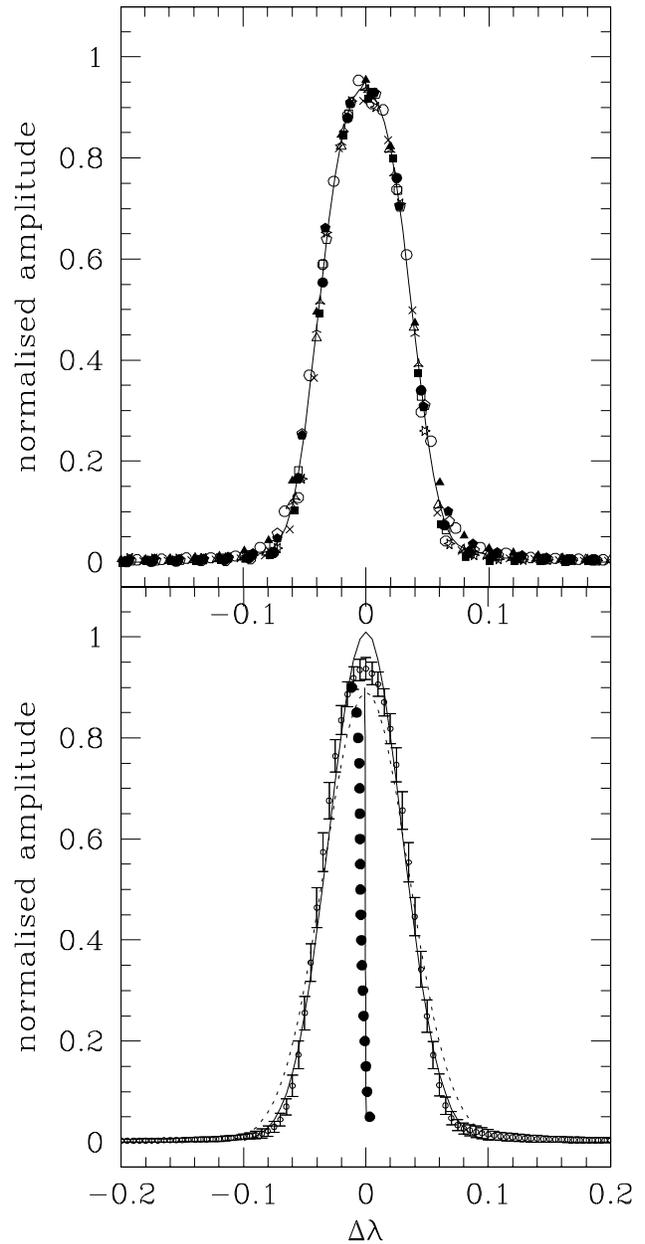}
%\centerline{\epsfxsize=084mm\epsfbox{li6hds_f1.eps}} 
\caption{
(a) {\it points}: Normalised flux versus distance from line centre for
13 ThAr lines. The peak intensity and central wavelength of each line
was determined from a Gaussian fit. {\it solid line}: numerical
instrumental profile obtained as the unweighted average of the points.
(b) {\it errorbars}: numerical instrumental profile and standard error
of each value. {\it solid curve}: Gaussian fit to instrumental profile
over $\pm$0.1 {\AA}. {\it dotted curve}: Gaussian fit to instrumental
profile over $\pm$0.2 {\AA}.  {\it vertical line}: line bisector of the
instrumental profile.  {\it filled dots}: line bisector enlarged
$\times$10 in wavelength.  }
\end{figure}

Although we use the numerical profile in our stellar analysis, it is
instructive to investigate analytic approximations. Standard
deviations were calculated over the intervals $\Delta\lambda$ =
$\pm0.1$ and $\pm0.2$ {\AA}, the former giving $\sigma$ = 0.030 {\AA} and
a tolerable fit to the line core, and the latter giving $\sigma$ =
0.037 {\AA} (Figure 1(b)). The larger standard deviation associated
with the wider interval arises because 1.5\% of the flux within
$\Delta\lambda$ = $\pm0.2$ falls between $\pm0.1$ and $\pm0.2$,
outside the 3$\sigma$ interval, whereas a true Gaussian would have
less than 0.3\% beyond 3$\sigma$. This inflates the standard deviation
if the Gaussian fit is evaluated out to $\pm$0.2 {\AA}. This shows that
the instrumental profile has broader wings than a Gaussian.  The
effect of (optionally) assuming a Gaussian instrumental profile is
addressed in the stellar analysis below.

Figure 1(b) also shows the bisector of the average instrumental
profile to illustrate its symmetry.  Without magnification one would
say it was perfectly symmetric. After magnification by a factor of
ten, a small s-shape can be seen through the profile, but for our
purposes the profile can be regarded as symmetric.

\subsection{\it Macroturbulence}

We sought to constrain the macroturbulence from stellar spectral lines
other than Li, and convolved the synthetic spectrum with a Gaussian
broadening function of FWHM $\Gamma_1$ km s$^{-1}$ to model
this. Synthetic spectra of five Ca I and Fe I lines (Table
\ref{turbulences}) were calculated iteratively to find values of
$A$(Ca) (or $A$(Fe)), $\Delta\lambda$ and $\Gamma_1$ that minimised
$\chi_r^2$.  As can be seen from column (6) of the Table, the values
of $\Gamma_1$ vary from line to line, but do not appear to correlate
with atomic number, wavelength, excitation potential or equivalent
width.  It was noted that the Ca I 6717$\mathrm{\AA}$ line has a weak
line in its long-wavelength wing that could not have been modelled
simultaneously to the same precision, potentially degrading our
results.  Ca~I 6717 also required a different $\Delta\lambda$ value,
$-8$~m{\AA} compared to $-18\pm2$~m{\AA} for the others.\footnote{Our
analysis of the Ca I line assumed a wavelength of 6717.688~\AA.  Dr
Bonifacio has kindly pointed out an unpublished measurement of the
wavelength by Rosberg \& Johansson, cited by Smith et al. (1998),
giving the value 6717.677~\AA, which is 11~m\AA\ less than we have
used.  Adopting the Rosberg \& Johansson value would change the
$\Delta\lambda$ value in Table~1 to $-19$~m\AA, which would be
consistent with the other values. However, it would not alter the
broadening values for the line, which also seem slightly anomalous,
possibly due to the presence of a weak blend as noted in the text.} It
was therefore decided to average the other four lines to give the
preferred macroturbulence, $\langle\Gamma_1\rangle = 4.65$ km
s$^{-1}$.  The consistency of the $\Delta\lambda$ values for the four
lines (excluding Ca~I 6717) argues for a systematic error in the
velocity correction applied during the data reduction as the source of
the shift.

We also investigated the result of adopting a radial-tangential
macroturbulent broadening profile, adopting the $\zeta_{\rm R}$ =
$\zeta_{\rm T}$ and $A_{\rm R}$ = $A_{\rm T}$ prescription of Gray
(1992, Chapter 18).  The Ca~I and Fe~I lines examined above
constrained $\zeta_{\rm RT}$ to 4.21$\pm$0.13~km~s$^{-1}$, whereupon a
value 4.20~km~s$^{-1}$ was adopted.

\begin{table*}
\begin{center}
\caption{Macroturbulence values for Ca and Fe lines \label{turbulences}}
\begin{tabular}{lcccccccc}
\hline
\hline
Ion 	& $\lambda$ & $\chi_{\rm lo}$	& W	&$\Delta\lambda$& $\Gamma_1$	& $\Gamma_{\rm c}$ & $\Gamma_2$ &$\zeta_{\rm RT}$\cr
 	&{\AA} 		& eV 		& m{\AA}	&m{\AA}		& km s$^{-1}$	& km s$^{-1}$	& km s$^{-1}$ & km s$^{-1}$\cr
(1) & (2) & (3) & (4) & (5) & (6) & (7)	&(8) &(9)\cr
\hline
Ca I & 6162.18 & 1.9 & 37.6 &$-$19& 4.83 & 6.03 & 5.14&4.40 \cr
Ca I & 6439.08 & 2.5 & 29.2 &$-19$& 4.60 & 5.75 & 5.10&4.10 \cr
Ca I & 6717.69 & 2.7 & 4.2  &$-$8 & 5.05 & 6.00 & 4.80&4.60 \cr
Fe I & 6494.99 & 2.4 & 26.7 &$-18$& 4.60 & 5.73 & 4.78&4.20 \cr
Fe I & 6678.00 & 2.7 & 12.4 &$-15$& 4.55 & 5.63 & 4.66&4.15 \cr
\hline
\multicolumn{4}{l}{average $\pm$ s.d. excl. 6717}&$-18\pm2$&4.65$\pm$0.13	&5.79$\pm$0.17	& 4.92$\pm$0.24 & 4.21$\pm$0.13\cr
\multicolumn{2}{l}{adopted}		  &&&&4.65			&5.80			&&4.20 \cr
\hline
Notes:\cr
\multicolumn{9}{l}{$\Gamma_1$ = FWHM of Gaussian macroturbulence adopting numerical instrumental profile.} \cr
\multicolumn{9}{l}{$\Gamma_{\rm c}$ = FWHM of composite Gaussian modelling macroturbulence and instrumental profile.} \cr
\multicolumn{9}{l}{$\Gamma_2$ = FWHM of Gaussian macroturbulence assuming $\Gamma_{\rm c}$ incorporates a Gaussian}\cr
&\multicolumn{8}{l}{instrumental profile with $\sigma$ = 0.030 {\AA}.} \cr
\hline
\end{tabular}
\end{center}
\end{table*}

To test the sensitivity of the results to the commonly-used assumption
of Gaussian profiles, the best fit for a single Gaussian broadening
function to model both the instrumental profile and macroturbulence
was also calculated.  The FWHM of this composite Gaussian is denoted
$\Gamma_{\rm c}$.  A similar spread in values for $\Gamma_{\rm c}$ as
for $\Gamma_1$ was observed. These values gave $\langle\Gamma_{\rm
c}\rangle = 5.80$ km s$^{-1}$. Since $\Gamma_{\rm c}$ represents a
convolution of the instrumental profile with the supposed Gaussian
macroturbulence $\Gamma_2$, column (8) in Table \ref{turbulences}
estimates the macroturbulent portion assuming the instrumental profile
is a Gaussian profile with a standard deviation of 0.03$\mathrm{\AA}$
(see \S~3.2).

\subsection{\it Lithium isotope ratio}

With the instrumental profile determined numerically and the
macroturbulence $\Gamma_1$ constrained from four Ca and Fe lines, the
synthesised and observed spectrum could be compared at Li 6707 {\AA}.
The $\chi^2$ statistic was calculated over the 35 pixels from 6707.50
{\AA} to 6708.20 {\AA}, where the Li profile is distinguishable from the
noise in the continuum. The continuum level was estimated out of this
wavelength range. The effects of the uncertainties in the
macroturbulence and the continuum level on the final results are
estimated in subsection 3.5.

The macroturbulence and the continuum level are two of the five free
parameters determined by the profile fitting to the Li line in the
analysis by Cayrel et al. (1999). However, the macroturbulence and the
continuum level are, at least in principle, able to be determined
independently of the Li line analysis. In contrast, the $^{6}$Li and
$^{7}$Li abundances must be determined from the Li line
analysis. Though the wavelength calibration and doppler correction can
be made from the detailed analysis of spectral lines other than the Li
line, there is still some uncertainty of the wavelengths of the line
transitions. For this reason, the wavelength zero-point is also
adjusted by the Li line analysis in the present work.

Calculations of $\chi_{\rm r}^{2}$ were performed covering a
reasonable range of isotope ratios, total Li abundances, and
wavelength shifts. These reduce the number of degrees of freedom by
three, to 32. It was found during early work using multiple sampling
of the independent variable $x$, that plots of $\chi_{\rm r}^{2}$ vs
$x$ could be fit very well by a quadratic function, and hence the
minimum value of $\chi_r^2$ and its corresponding $x$ value could be
found after a few calculations.  This yielded
$^6\mathrm{Li}/^7\mathrm{Li}$ = 0.001 as the best fit (see Table
\ref{chis} and corresponding $\chi_{\rm r}^2$-plots in Figure~2).

%: $^6\mathrm{Li}/^7\mathrm{Li}$ = 0, 0.02, 0.04 and
%0.06; log (Li)=xx, xx and xx, and $\Delta \lambda$=xx, xx, and
%xx. (SEAN, CAN YOU GIVE THESE VALUES?)

\begin{table*}
\begin{center}
\caption{Isotope ratios inferred from different spectral models\label{chis}}
\begin{tabular}{lcccccc}
\hline
\hline
Inst. & Macroturb.  & Macrot. & $^6$Li/$^7$Li & $\chi_r^2$ & A(Li) & $\delta \lambda$ \cr
profile & constraint & km~s$^{-1}$ &  & & & m{\AA} \cr
(1) & (2) & (3) & (4) & (5) & (6) & (7) \cr
\hline
\multicolumn{5}{l}{Preferred solution}\cr
HDS     & Ca $\&$ Fe & $\Gamma_1$=4.65 	    & 0.001$^{+0.025}_{-0.001}$	& 0.74 & 2.184 & $-28.4$ \cr
\multicolumn{5}{l}{Other explored solutions}\cr
HDS     & Li 	     & $\Gamma_1$=4.53	    & 0.005 			& 0.71 & 2.181 & $-28.0$\cr
HDS	& Ca $\&$ Fe & $\zeta_{\rm RT}$=4.20& $-0.007$			& 0.83 & 2.188 & $-29.2$\cr
Gauss. 	& Ca $\&$ Fe & $\Gamma_{\rm c}$=5.80& 0.002                    	& 0.81 & 2.183 & $-28.3$\cr
\hline
\end{tabular}
\end{center}
\end{table*}

The procedure was repeated for the pure Gaussian instrumental profile,
giving essentially the same result, $^6\mathrm{Li}/^7\mathrm{Li}$ =
0.002.  It can be seen from Figure~2 that adopting the pure Gaussian
profile for the combined macroturbulent and instrumental profiles is
in close agreement with the preferred result based on the numerical
instrumental profile. The calculation using the numerical instrumental
profile also gives a marginally smaller $\chi_r^2$ value.

\begin{figure}
% tife_f2.eps goes here
% \vspace{11cm}
\epsfxsize=084mm
\epsfbox[44 173 285 422]{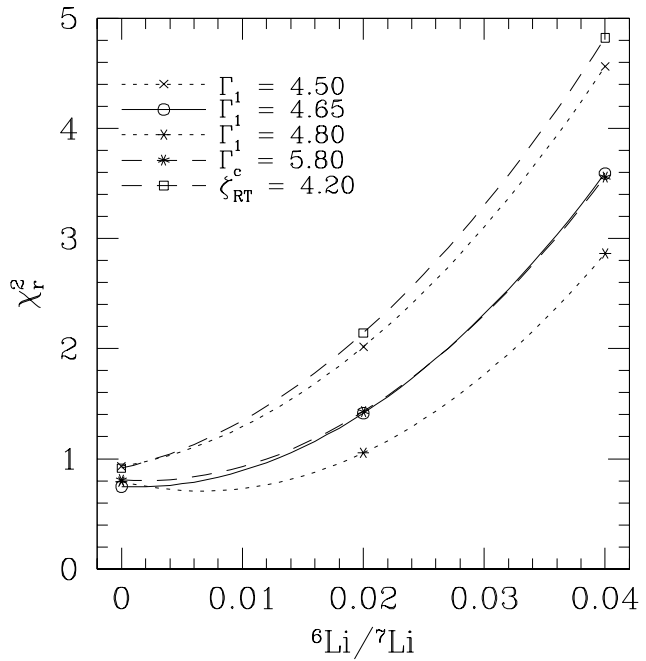}
\caption{
$\chi_{\rm r}^2$ values for five spectrum synthesis models 
as a function of input lithium isotope ratio. Points indicate the actual calculations, while curves give the fitted quadratic functions used to identify the minimum.
{\it solid curve}: preferred solution using numerical instrumental profile and 
Gaussian macroturbulence $\Gamma_1$ = 4.65~km~s$^{-1}$
constrained from four Ca and Fe lines.
{\it dotted curves}: solutions in which $\Gamma_1$ was varied to $\pm$1$\sigma$
from the mean, to 4.50 and 4.80~km~s$^{-1}$.
{\it dashed curves}: Alternative solutions in which 
(i) pure Gaussian functions were used for instrumental and macroturbulent 
broadening ($\Gamma_{\rm c}$ = 5.80~km~s$^{-1}$) and
(ii) the HDS instrumental profile and radial-tangential macroturbulence
($\zeta_{\rm RT}$ = 4.20~km~s$^{-1}$) were assumed.
}
\end{figure}

To test whether the Ca and Fe lines gave a reasonable value of
$\Gamma_1$ to use on the lithium lines, we performed a series of
iterations on $A$(Li), $\Delta\lambda_{6707}$ and $^6$Li/$^7$Li in
which $\Gamma_1$ was also allowed to vary.  The $\chi_r^2$ minimum was
found for $\Gamma_1$ = 4.53 km s$^{-1}$, shown in Table \ref{chis}.
This lies within 1$\sigma$ of the preferred value,
4.65~km~s$^{-1}$. Allowing $\Gamma_1$ to vary used up another degree
of freedom, but nevertheless put $\chi_r^2$ marginally below the value
for when $\Gamma_1$ took the preferred value constrained by the Ca and
Fe lines.

The lower broadening value weakens the wings of the profile and hence
requires a higher $^6$Li/$^7$Li ratio to fit the specific observation,
but $^6$Li/$^7$Li remained below 0.01.  Even allowing
$\Delta\lambda_{6707}$ and $A$(Li) to vary, we found an almost linear
relation between the $^6$Li/$^7$Li ratio inferred and the adopted
$\Gamma_1$, $^6$Li/$^7$Li~=~$-$0.038$\Gamma_1$~+~0.1777 over the
interval 4.50 $<$ $\Gamma_1$ $<$ 4.80; probably this relation holds
over a much wider range of macroturbulent values.

Figure~3 compares the observational data to the best fitting model and
related models having $^6$Li/$^7$Li = 0.02 and 0.04.

We note that the equivalent width of the Li absorption line determined
by the best-fit synthetic spectrum is 47.8~m{\AA}. This value agrees
well with that derived by Ford et al. (2002) for the same spectrum
(48.1~m{\AA}). We also note that, even though the Li abundances
derived by the present analysis are given in Table 2, the values are
dependent on the choice of effective temperatures, while the isotope
ratio is insensitive to the atmospheric parameters.

\begin{figure}
% lihds_f3.eps goes here
% \vspace{11cm}
\epsfxsize=084mm
\epsfbox[18 327 323 689]{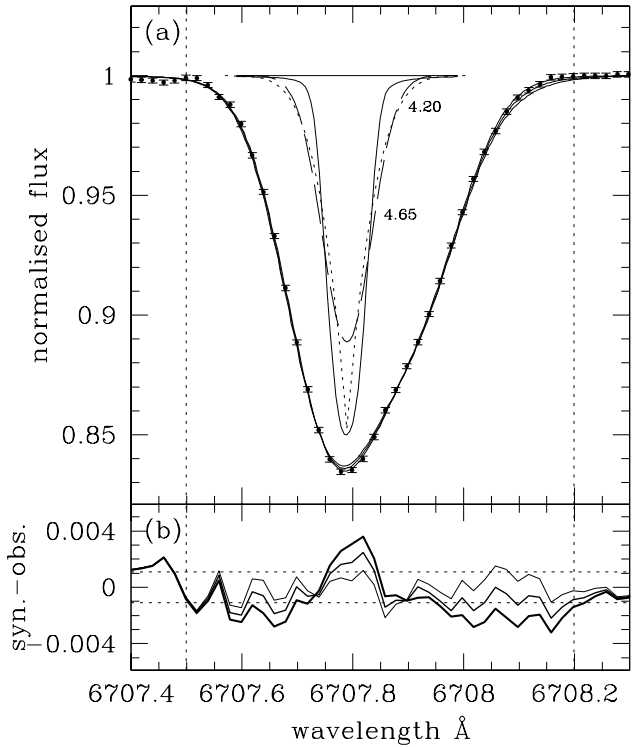}
\caption{
(a) Comparison of ({\it curves}) synthetic and ({\it filled circles})
observed Li-doublet spectra 
for the preferred, best-fitting model in Figure~2 having $^6$Li/$^7$Li = 0.00
and also models at 0.02 and 0.04. Note that the higher $^6$Li/$^7$Li ratios
demand lower $^7$Li abundances, resulting in a weakening of the line core.
The inset shows three equal-area broadening profiles for 
({\it solid curve}) the HDS instrumental profile,
({\it dashed curve}) $\Gamma_1$ = 4.65~km~s$^{-1}$, and 
({\it dotted curve}) $\zeta_{\rm RT}$ = 4.20~km~s$^{-1}$.
The vertical dotted lines indicate the wavelength region over which the Li
$\chi^2$ analysis was conducted.
(b) Difference in normalised flux between synthetic and observed spectra.
Light, medium and heavy lines correspond to $^6$Li/$^7$Li = 
0.00, 0.02 and 0.04 respectively.
The horizontal dotted lines correspond to $\pm$1$\sigma$ in the observed flux.
}
\end{figure}

\subsection{Uncertainties}

We first estimate the statistical errors inherent in the spectrum with
S/N$\simeq$1000. We have tried to quantify this in two ways.

The first method is to consider the interpretation of the $\chi_r^2$
statistic.  The probability that a $\chi^2$ value as large as that
measured should occur by chance is tabulated in many statistics
books. Our plot of $\chi_r^2$ vs $^6$Li/$^7$Li for the preferred
analysis method indicates that very low $\chi_r^2$ values were
obtained at the best fitting isotope ratio, corresponding to high
probabilities that random statistical fluctuations could produce
$\chi^2$ values this large. The random fluctuations modelled in the
$\chi^2$ formalism are, we recall, the $\sigma$ values associated with
the noise in the continuum, in our case 0.0011. Away from the minimum,
$\chi_r^2$ is larger and the probability that random fluctuations
could produce such large $\chi^2$ values decreases.  For 32 degrees of
freedom, the probability falls to 84.1\% at $\chi_r^2$ = 1.325, and to
97.8\% at $\chi_r^2$ = 1.633. These correspond to $^6$Li/$^7$Li =
0.019 and 0.023, respectively, for the adopted value of $\Gamma$ =
4.65~km~s$^{-1}$.

However, due to the rebinning of the original spectra that occurs
during data reduction, photon errors in adjacent pixels are not fully
independent. As the $\chi^{2}$ test assumes independent errors in
adjacent pixels, the calculated chi-squared value is only an
approximation to the value which ought to be minimised, as pointed out
by Cayrel et al. (1999) and Bonifacio \& Caffau (2003). We continue to
use the $\chi^{2}$ procedure and in particular select as our preferred
parameter set that minimises the calculated $\chi_{\rm r}^{2}$
value, but we are unable to associate a confidence interval to any
particular chi-squared values that we calculate.

The second method involves conducting a Monte Carlo test in which a
synthetic spectrum corresponding closely to the observed profile is
subjected to Gaussian errors having the same distribution ($\sigma$ =
0.0011) as the real data. A series of noisy, synthesised spectra based
on a single input model are then subjected to the same
$\chi^2$-fitting procedures as the real analysis, to determine whether
the input parameters can be recovered and with what accuracy. We
performed 300 simulations to check the size of likely errors.  Note
that we are primarily interested in the spread of the results rather
than the central value.  For a synthetic spectrum having $A$(Li) =
2.223, a wavelength error $\Delta\lambda$ = 0.001 {\AA}, and
$^6$Li/$^7$Li = 0.014, we inferred the following parameters: $A$(Li) =
2.224$\pm$0.001, $\Delta\lambda$ = 0.0012$\pm$0.0011, and
$^6$Li/$^7$Li = 0.011$\pm$0.004. The inferred isotope ratios deviate
from the input value with an RMS of 0.0042. This test suggests that,
where the macroturbulence is constrained from other spectral features,
the $^6$Li/$^7$Li ratio can be recovered from data with S/N=900 at an
RMS deviation = 0.0042. If the distribution of the results of the
above simulation is assumed to be the Gaussian, the 3$\sigma$ limit is
0.013 in the $^6$Li/$^7$Li ratio.

The Monte Carlo estimate of the uncertainty in $^6$Li/$^7$Li provides
a smaller error than that inferred from the $\chi^2$ statistic. Since
the error estimate from the $\chi^2$ approach contains a difficulty
in our case as mentioned above, we adopt the value estimated by the
Monte Carlo simulation (0.013 in the $^6$Li/$^7$Li ratio) as the
statistical error. 

It is easy to calculate the 
{\bf
effect
}
of uncertain $\Gamma$ values in the
inferred $^6$Li/$^7$Li ratio because, as noted above, there is a
linear dependence of the isotope ratio on the assumed value. A
characteristic uncertainty $\sigma_\Gamma$ = 0.15 km s$^{-1}$
translates to 0.006 in the isotope ratio. 

We also estimated the errors due to the uncertainty of continuum
normalization by changing the continuum level in the analysis. We
found almost a linear correlation between the assumed continuum level
and resulting $^6$Li/$^7$Li ratio. The assumption of by 0.1\% higher
continuum results in by $-0.010$ lower $^6$Li/$^7$Li ratio.

Combining the limit ($^6$Li/$^7$Li $<$0.013) by the statistical error
in quadrature with errors due to the uncertainties of macroturbulence
and continuum level, we infer the upper limit of $^6$Li/$^7$Li to be
0.018.

\section{Discussion }

The very low $^6$Li/$^7$Li ratio found for HD~140283 is well below the
value 0.040$\pm$0.015 inferred by S.G.R. from an earlier analysis of
poorer data (Deliyannis \& Ryan 2000). S.G.R's previous analysis
differed in several respects: the S/N was lower (435), and the
sampling coarser (0.026 {\AA} pix$^{-1}$), for the same resolving
power. The reduction procedure also differed: in that work, an effort
was made to constrain the fitting parameters independently rather than
through a highly iterative $\chi^2$ analysis as employed here. It is
probable that the present procedure has resulted in better
optimisation of the fitting parameters than was achieved in the
previous work.  Application of the current technique to the older
spectrum gave $^6$Li/$^7$Li = 0.011 when $\Gamma_{\rm c}$ =
5.90~km~s$^{-1}$ was constrained by the Ca~I 6162~{\AA} line, and
0.034 when $\Gamma_{\rm c}$ = 5.38~km~s$^{-1}$ was constrained by the
Li~I 6707~{\AA} fit. These large variations in the inferred isotope
ratio emphasise the need for good constraints on the macroturbulence
if a reliable isotope fraction is to be obtained.

The analysis presented above is based solely on 1D model atmospheres.
Analyses using 3D model atmospheres, which have begun to appear in
recent years (e.g. Asplund et al. 1999), offer the hope of replacing
the empirical micro- and macroturbulent parameters with
physically-based velocity structures.  The importance of this is clear
from the strong correlation between the input (adopted) macroturbulent
broadening and the output isotope ratio.  It can only be hoped that 3D
calculations will soon become commonly accessible to all stellar
spectroscopists.

Deliyannis' suggestion that sufficiently warm subgiants might preserve
their main-sequence $^6$Li complement finds no support from the
current observation, but nor is the suggestion necessarily
defeated. At $T_{\rm eff}$ = 5750 K, HD~140283 is very close to the
effective temperature at which even $^7$Li is seen to be diluted in
subgiants (Pilachowski, Sneden \& Booth 1993), so it is possible that
$^6$Li has begun to be depleted in this object. It is also amongst the
most metal-poor of the stars in which $^6$Li has been sought, so
Galactic production may have been lower at this epoch. (Bear in mind,
nevertheless, that $^6$Li production via cosmic ray fusion is not as
metallicity-dependent as production via spallation.) For these reasons
it is still appropriate to examine warmer and/or higher-metallicity
subgiants in an effort to detect $^6$Li in them.

% \section*{Theoretical Discussion}

However, it is also possible that our upper limit reflects a genuinely
low $^6$Li abundance at production.  On the premise that depletion has
not been significant, and that the HD140283 result is representative
of stars in its metallicity range, we consider below the implications
for different production models of Pop. II $^6$Li.  Fig. 4 shows our
data together with previously published results of $^6$Li detections
in HD 84937, BD+26$^\circ$3578 and G271-162 (Smith et al. 1998; Nissen
et al. 2000 and references therein).  Also plotted are tentative
detections for three additional stars from Asplund et al. (2001),
CD-30$^\circ$18140, G13-9 and HD160617, achieved through preliminary
analysis of recent VLT/UVES observations.  Note that although HD140283
was additionally reported in Asplund et al. (2001) as a possible
detection, their latest analysis for this star is consistent with our
limit here.  Taken at face value, comparison of our upper limit with
the other detections may suggest a relatively steep increase of
$^6$Li/H with metallicity near [Fe/H] $\simeq -2.4$, followed by a
slower rise.  Alternatively, the data set may indicate typical $^6$Li
abundances that are a factor of 2--3 lower than the highest measured
values, which may be consistent with previous upper limits for a few
other stars (Smith et al. 1998, Hobbs et al. 1999).

The most widely discussed models so far for Pop. II $^6$Li synthesis
are based on spallation and/or fusion reactions induced by cosmic rays
originating from supernovae (SNe).  While they can successfully
explain the Be and B observed in Pop. II stars, accounting for the
observed $^6$Li is rather problematic; they require either an
implausibly high cosmic ray injection efficiency (Ramaty et al. 2000,
Suzuki \& Yoshii 2001), or the presence of an additional low energy
cosmic ray component lacking observational support (Vangioni-Flam et
al. 1999).  Even if the typical values of $^6$Li abundances turn out
to be lower than the highest measurements by a factor of 2--3, the
data would still significantly exceed conservative predictions of such
models assuming standard SN CR energetics and spectra, drawn as a
dashed line (D) in Fig. 4 (see Suzuki \& Inoue 2002).  In some
secondary SN CR models, a steeply rising log($^6$Li/H)-[Fe/H] relation
can occur depending on the uncertain O-Fe relation (Fields \& Olive
1999), but this is highly unlikely at the observed abundance levels in
this metallicity range.

A very different scenario for $^6$Li production has recently been put
forth by Suzuki \& Inoue (2002): nuclear reactions induced by cosmic
rays accelerated at structure formation (SF) shocks,
i.e. gravitational virialization shocks driven by the infall and
merging of gas in sub-Galactic haloes during hierarchical build-up of
structure in the early Galaxy.  Such shocks are inevitable
consequences of the currently standard theory of structure formation
in the universe.  Estimates for the specific energy dissipated at the
main SF shock accompanying the final major merger give $\epsilon_{SF}
\simeq 0.4 {\rm keV}$ per particle, compared to that for early SNe at
$\epsilon_{SN} \sim 0.15 {\rm keV}$ per particle.  Thus SF shocks can
be potentially more energetic than SNe at early Galactic epochs, and
the associated CRs can explain the $^6$Li observations more naturally.
Since such shocks do not eject freshly synthesized CNO nor Fe,
$\alpha-\alpha$ fusion is the dominant production channel at low
metallicities, which can generate large amounts of $^6$Li with little
Be or B and no direct correlation with Fe.  A unique evolutionary
behavior can arise, whereby $^6$Li increases quickly at low
metallicity (reflecting the main epoch of Galactic SF), followed by a
plateau or a slow rise, in marked contrast to SN CR models for which
$\log$ ($^6$Li/H) vs. [Fe/H] can never be much flatter than linear.  Shown
in Fig. 4 are three possible model curves for different parameter
values of $t_{SF}$, the main epoch of Galactic SF relative to halo
chemical evolution, $\tau_{SF}$, the main duration of SF, and
$\gamma_{SF}$, the spectral index of injected particles (see Suzuki \&
Inoue 2002 for more details).

The current data set including the HD140283 upper limit may be
consistent with curve C, corresponding to strong shocks with hard CR
spectra (expected in case the preshock gas is efficiently cooled by
radiation), or a total SF CR energy a factor of $\sim$ 2.7 below the
estimates described above (implying a lower total halo mass after the
merger).  If the steep rise of $^6$Li is real, curve B may be a better
representation, where the main SF shock at the final major merger
occurs near [Fe/H] $\sim$ -2, possibly being compatible with some
other lines of evidence (e.g. Chiba \& Beers 2000).  
% For conclusive
% tests of the SF shock picture, more detailed and less parameterized
% modeling using e.g. semi-analytic techniques is necessary (Suzuki,
% Nagashima \& Inoue, in preparation).  
For conclusive tests of the SF shock picture, more detailed
and less parameterized modelling utilizing e.g numerical simulations
of galaxy formation is necessary.
A further crucial prediction of
the scenario is correlations between the $^6$Li abundance and the
kinematic properties of Pop. II stars, which may also provide
interesting insight into how the Galaxy formed.

The low upper limit for HD140283 is an important addition to the
observational database for $^6$Li in Pop. II stars.  However, it is
still insufficient for discriminating between different production
models, especially in view of the uncertain extent of depletion.  To
elucidate the true origin of $^6$Li as well as the effects of stellar
depletion in Pop. II stars, it is essential to obtain high quality data
for a larger sample of stars with a wide range of metallicities and
higher surface temperatures, an important goal for instruments such as
the Subaru HDS.

\begin{figure}
% lihds_f4.eps goes here
% \vspace{11cm}
\epsfxsize=084mm
\epsfbox[80 371 538 693]{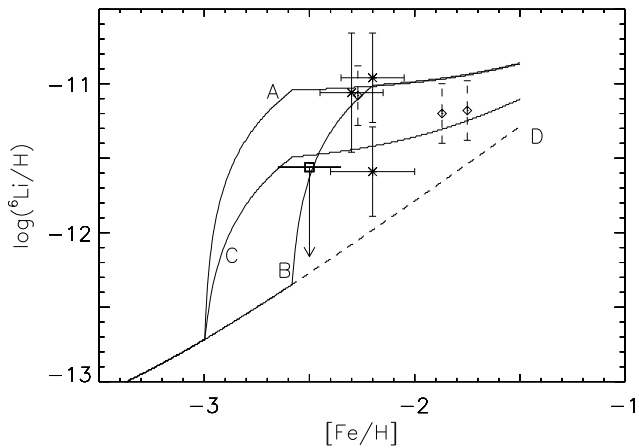}
\caption{Current observational data for $^6$Li/H vs. [Fe/H] in Pop. II
stars, compared with different models.  The square, crosses and
triangles correspond to the Subaru HDS upper limit for HD140283,
previous positive detections, and tentative VLT/UVES detections of
Asplund et al. (2001), respectively.  Model curves show the mean
abundance predictions for SN CRs only (D, dashed), and SN plus SF CRs
(A,B,C).  Labels correspond to the following sets of parameters for
$t_{SF}$ [Gyr], $\tau_{SF}$ [Gyr] and $\gamma_{SF}$: A (0.22, 0.1, 3),
B (0.32, 0.1, 3) and C (0.22, 0.1, 2).  }

\end{figure}

\section{Conclusions} We have obtained a high-S/N (900-1100),
high-resolving-power (R=95000) spectrum of the metal-poor subgiant
HD~140283 using HDS on the 8.2~m Subaru telescope, in an effort to
measure its $^6$Li/$^7$Li isotope ratio.  We computed synthetic
spectra using a 1-D model-atmosphere code, and used a $\chi^2$ test to
find the best fitting parameters.  We find an isotope ratio consistent
with zero, $^6$Li/$^7$Li = 0.001, with an upper limit
of $^6$Li/$^7$Li$<$0.018 estimated by a Monte Carlo analysis.  

%We regard this limit as conservative. A Monte
%Carlo analysis in which noisy synthesised spectra were subjected to
%the same fitting process suggests that a more restrictive result could
%be stated, as it indicates that the RMS error on measurements of
%$^6$Li/$^7$Li is only 0.0044.

This measurement supersedes an earlier detection
(0.040$\pm$0.015($1\sigma$)) by one of the authors (Deliyannis \& Ryan
2000). HD~140283 provides no support for the suggestion that Pop. II
stars may preserve their $^6$Li on the portion of the subgiant branch
where $^7$Li is preserved. However, this star does not defeat the
suggestion either; being at the cool end of subgiant branch of the
Spite plateau, it may be sufficiently cool that $^6$Li depletion has
already set in, or the star may be sufficiently metal poor that little
Galactic production of $^6$Li had occurred. Continued investigation of
other subgiants is necessary to test this idea.

We also consider the HD140283 upper limit along with other
measurements in the context of $^6$Li production models, particularly
the structure formation shock scenario, in which unique evolutionary
trends can be expected without any direct relation to the abundances
of Be, B, CNO or Fe.  If HD140283 has not been depleted in $^6$Li, the
upper limit can impose interesting constraints on the epoch and
efficiency of dissipative dynamical processes that occurred during the
formation of the Galaxy.  Further observations together with improved
theoretical modeling should allow us to test the structure formation
scenario more critically and quantitatively, and to assess the value
of $^6$Li as a dynamical probe of the early Galaxy.

\section*{Acknowledgements}

S.G.R. acknowledges numerous discussion with Dr C. P. Deliyannis on
lithium processing in stars.  This work was supported financially by
PPARC (PPA/O/S/1998/00658) and the Nuffield Foundation
(NUF-URB02). The high quality spectrum of HD~140283 used in the
present study was obtained in a commissioning run of the HDS
instrument group.
S.I. acknowledges financial support from the 
Foundation for Promotion of Astronomy.

% \vfill
% \eject

\label{lastpage}

\end{document}